\documentclass[a4paper,fleqn]{mnras}

\usepackage{mathptmx}

\usepackage{txfonts}

\usepackage{graphicx}	

\title[Dynamical differentiation
  between W3(H$_2$O) and W3(OH)]{SMA observations of the W3(OH) complex: Dynamical differentiation between W3(H$_2$O) and W3(OH)}

\author[S. L. Qin et al.]{
Sheng-Li Qin,$^{1}$\thanks{E-mail: slqin@bao.ac.cn}
Peter Schilke,$^{2}$
Jingwen Wu,$^{3}$
Tie Liu,$^{4}$
 Yuefang Wu,$^{5}$
\newauthor
\'Alvaro S\'anchez-Monge,$^{2}$
and Ying Liu$^{6}$
\\
$^{1}$Department of Astronomy, Yunnan University, and Key Laboratory of Astroparticle Physics of Yunnan Province,
            Kunming, 650091, China \\
$^{2}$I. Physikalisches Institut, Universit\"at zu K\"oln,
             Z\"ulpicher Str. 77, 50937 K\"oln, Germany  \\
$^{3}$Department of Physics and Astronomy, University of California, Los
             Angeles, CA 90095, USA \\
$^{4}$Korea Astronomy and Space Science Institute 776, Daedeokdaero,
Yuseong-gu, Daejeon, Republic of Korea 305-348 \\
$^{5}$Department of Astronomy, Peking University, Beijing, 100871, China \\
$^{6}$Department of Physics and Hebei Advanced Thin Film Laboratory, Hebei Normal University, Shijiazhuang 050024, China
}

\date{Accepted XXX. Received YYY; in original form ZZZ}

\pubyear{2015}
\begin{document}
\label{firstpage}
\pagerange{\pageref{firstpage}--\pageref{lastpage}}
\maketitle

%

\begin{abstract}

We present Submillimeter Array (SMA) observations of the HCN\,(3--2) and
HCO$^+$\,(3--2) molecular lines toward the W3(H$_2$O) and W3(OH) star-forming
complexes. Infall and outflow motions in the W3(H$_2$O) have been
characterized by observing HCN and HCO$^+$ transitions. High-velocity
blue/red-shifted emission, tracing the outflow, show multiple knots, which
might originate in episodic and precessing outflows. `Blue-peaked' line
profiles indicate that gas is infalling onto the W3(H$_2$O) dust core. The
measured large mass accretion rate,
2.3$\times$10$^{-3}$~M$_{\odot}$~yr$^{-1}$, together with the small
free-fall time scale, 5$\times$10$^{3}$~yr, suggest W3(H$_2$O) is in an early
evolutionary stage of the process of formation of high-mass stars. For the
W3(OH), a two-layer model fit to the HCN and HCO$^+$ spectral lines and
Spizter/IRAC images support that the W3(OH) H{\sc ii} region is expanding and
interacting with the ambient gas, with the shocked neutral gas  being
  expanding with an expansion timescale of 6.4$\times$10$^{3}$~yr. The
observations suggest different kinematical timescales and dynamical states for the W3(H$_2$O) and W3(OH).
\end{abstract}


%

\section{Introduction}
     
Current theories and observations suggest that low-mass stars form by
gravitational collapse and subsequent accretion through a circumstellar disk,
with the accretion energy driving a collimated outflow perpendicular to the
plane of the disk. Therefore, accretion disks and outflows are physically
connected and are key ingredients in the process of formation of low-mass
stars. For high-mass stars, their formation process is still not well
understood. Outflows are frequently detected in high-mass star forming
regions, but it is thought that high-mass star formation is not merely a
scaled-up version of the low-mass star formation process characterized by
larger outflow parameters (e.~g., Wu et al.\ 2004), poor collimation around
massive young stars (e.~g., Shepherd \& Churchwell 1996; Zhang et al. 2005) and higher accretion rates (see review by Zinnecker \& Yorke
2007). Additionally, disks  in high-mass young stellar objects are still
elusive --- only a few of them have been detected (Zhang et al.\ 1998;
 Shepherd, Claussen \& Kurtz 2001; Patel et al.\ 2005;
Jiang et al.\ 2005; S\'anchez-Monge et al.\ 2013a, 2014) --- probably due to their short lifetimes, the necessity of
achieving high angular resolution (not available in the observations prior to
ALMA) and source confusion in dense clusters. Therefore, the study and
characterization of the outflow and infall processes involved in the formation
of high-mass stars need to be  understood in more detail. Since high-mass
stars form in dense cores, dense gas tracers such as HCN are useful in
identifying outflow and infall motions (e.~g., Liu et al.\ 2011a, 2011b, 2013,
2015; Wu et al.\ 2005). It has also been demonstrated that the profile of
spectral lines such as HCN (3--2) and HCO$^{+}$ (3--2) are good tools to
search for and study the infall motions (see simulations by Smith et
al.\ 2012, 2013; and Chira et al.\ 2014).

Located at a distance of 2.04~kpc (Hachisuka et al.\ 2006), the W3(OH) complex, consisting of two objects: W3(OH) itself and W3(H$_2$O), is one of the well-studied high-mass star forming regions. W3(OH) is associated with an ultracompact (UC) H{\sc ii} region powered by an O7 star ( Dreher \& Welch 1981; Wilner, Welch \&  Forster 1995), while W3(H$_2$O) is rich in H$_2$O maser emission and organic molecules (Alcolea et al.\ 1993; Wyrowski et al.\ 1999; Hern\'andez-Hern\'andez et al.\ 2014; Qin et al.\ 2015), suggesting that it is at an earlier evolutionary stage. The observations suggest that W3(H$_2$O) has a mass of 10--20~M$_{\odot}$ and a luminosity of 10$^3$--10$^4$~L$_{\odot}$ (Wilner, Welch \& Forster 1995).  High-angular resolution observations of the continuum and CH$_3$CN lines resolve W3(H$_2$O) into three subcomponents: A, B, and C; with A and C likely forming a binary system (Wyrowski et al.\ 1999; Chen et al.\ 2006). Non-thermal radio jets and water maser outflows have been revealed with VLA and VLBA (Alcolea et al.\ 1993; Reid et al.\ 1995; Wilner et al.\ 1999) observations along the east-west direction centered on W3(H$_2$O). Two CO molecular outflows (with PA=40$^{\circ}$ and 15$^{\circ}$) and probably a rotating disk, have been studied in the W3(H$_2$O) with interferometers (Zapata et al.\ 2011). Single-pointing spectral line surveys reveal infall motions in the W3(OH) complex (e.~g., Wu \& Evans 2003; Wu et al.\ 2010), but the infalling center and the gas accretion rate have not been well determined.  The kinematics of W3(OH) are still under debate. OH masers and NH$_3$ absorption features suggest a collapsing outer envelope (Reid et al.\ 1980; Garay, Reid \& Moran 1985; Guilloteau, Stier \& Downes 1983), while expansion motions were proposed based on OH maser observations and multi-epoch VLA observations (Bloemhof, Reid \& Moran 1992; Kawamura \& Masson 1998).

In this paper, we present HCN and HCO$^{+}$ line observations of the W3(OH)
star-forming complex obtained with the Submillimeter Array (SMA)\footnote{The
  Submillimeter Array is a joint project between the Smithsonian Astrophysical
  Observatory and the Academia Sinica Institute of Astronomy and Astrophysics
  and is funded by the Smithsonian Institution and the Academia Sinica.}  and
Caltech Submillimeter Observatory (CSO). The results show kinematical
differences between the W3(H$_2$O) and W3(OH), in which infall motion seems to
occur only towards the W3(H$_2$O) with evidence for jet procession, while gas
expansion is observed towards W3(OH).

\section{Observations}

The SMA was used to observe the HCN\,(3--2) and HCO$^{+}$\,(3--2) transitions at 267~GHz. We summarize here the main aspects of the observations, while further details and the data reduction process can be found in Qin et al.\ (2015). The observations cover the HCN\,(3--2) and HCO$^{+}$\,(3--2) transitions at 267~GHz, with an spectral resolution of 0.406~MHz (corresponding to $\sim$0.5~km~s$^{-1}$). The weather conditions were good during the observations resulting in system temperatures of around 200~K. Flux calibration was performed using Uranus and we estimate an uncertainty of 20\%. The rms (1~$\sigma$) noise level is 0.15~Jy~beam$^{-1}$.

Single-dish observations of HCN (3--2) were obtained using the 10.4~m
telescope at the Caltech Submillimeter Observatory (CSO), on 2003, January
15. We covered an area of 110$^{\prime\prime}\times110^{\prime\prime}$, with
10$^{\prime\prime}$ spacing. The beam size and main beam efficiency are
28$^{\prime\prime}$ and 60\%, respectively. These observations were used to
complement the missing short spacing information of the HCN\,(3--2) SMA
observations. The CSO data were converted to Miriad format. The task MOSMEM
was used to perform a joint deconvolution of the interferometric and
single-dish data. The resulting combined dataset has a spatial resolution of
$\sim$2$^{\prime\prime}.3$$\times$2$^{\prime\prime}.1$ (PA=--60$^{\circ}$)
with robust weighting, and a sensitivity of 0.17~Jy~beam$^{-1}$, with
1~Jy~beam$^{-1}$ corresponding to a main beam brightness temperature of
$\sim$3.5~K.

\section{Results}
\subsection{\emph{Gas Distribution}}
\subsubsection{HCN spectra obtained with the CSO}

\begin{figure*}
 \includegraphics[width=12.81cm,angle=-90]{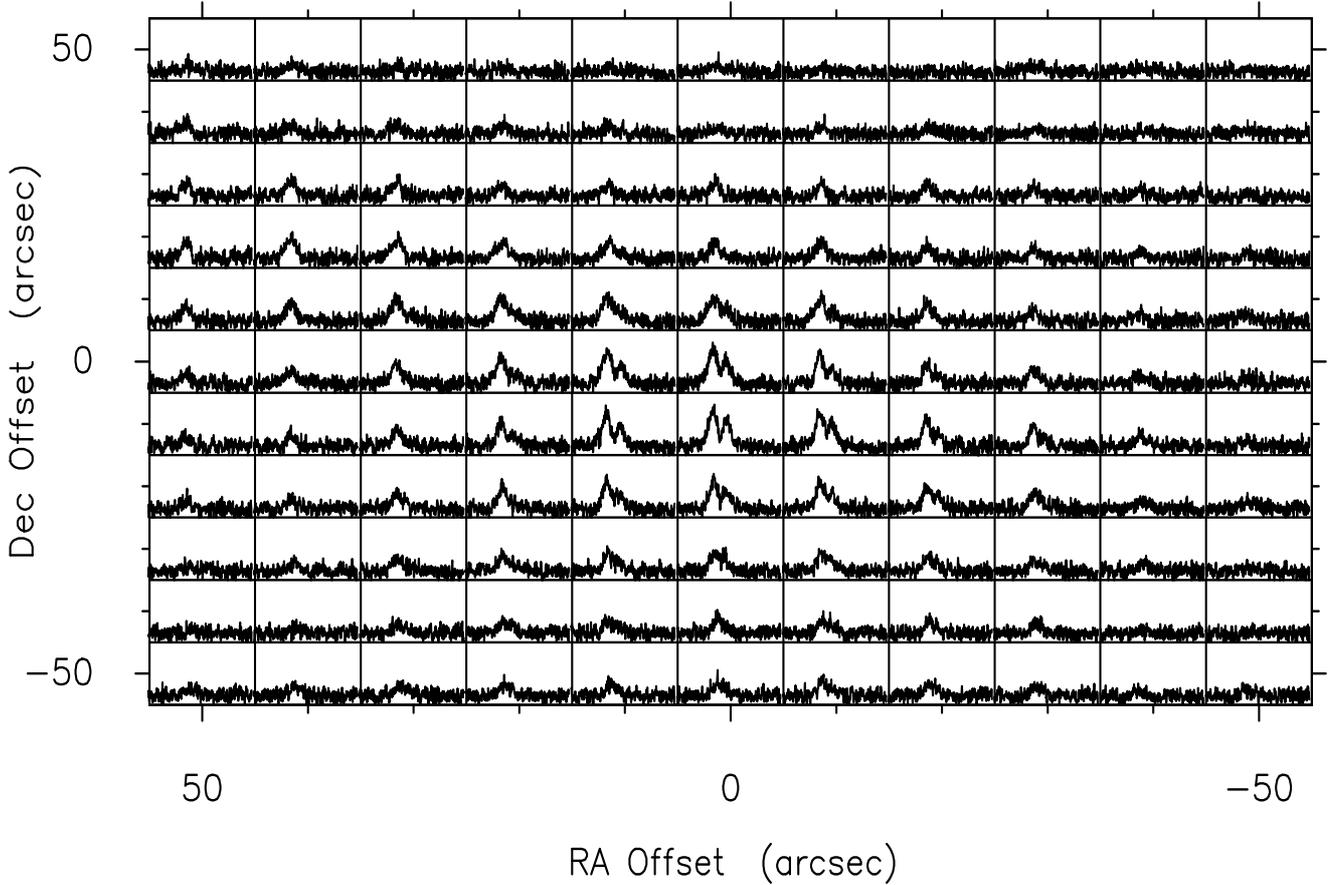}
 \caption{CSO HCN\,(3--2) spectral line observations. The mapped area covers an extension of 110$^{\prime\prime}$$\times$110$^{\prime\prime}$, with 10$^{\prime\prime}$ spacing. The beam size of the telescope at 267~GHz is $\sim$28$^{\prime\prime}$. }\label{}
\end{figure*}

The W3(H$_2$O) hot core was first studied in the HCN\,(1--0) line at 89~GHz by
Turner \& Welch (1984). However, their observations lack a good spectral
resolution (4.1~km~s$^{-1}$), and can not resolve detailed kinematics.  The
single pointing CSO observations of W3(OH) by Wu \& Evans (2003) show that the
HCN\,(3--2) line has a double peak, with the blue-shifted peak being stronger
than the red-shifted one (hereafter `blue profile'), while the
H$^{13}$CN\,(3--2) line peaks at the absorption dip of the HCN\,(3--2), which was used by the authors to identify W3(OH) as a collapsing candidate object. However, their low-spatial resolution observations can not determine the center of collapse. Our CSO observations of the HCN\,(3--2) line, covering an area of 110$^{\prime\prime}\times110^{\prime\prime}$, are shown in Figure~1. The `blue-profile' is not only seen towards the central position but towards a large area, suggesting that the W3(OH) complex is overall undergoing collapse, similar to what is observed in the G10.6$-$0.4 star forming region (Liu et al.\ 2013).

\subsubsection{Channel maps and velocity field}

Figure~2 shows the velocity channel map of the HCN\,(3--2) line at 265.886~GHz
as observed with the SMA (green contours). The combined SMA and CSO data
(SMA+CSO, red contours) are overlaid. The structure of the emission is the
same in the SMA and SMA+CSO datasets in the velocity ranges from $-$69 to
$-$53~km~s$^{-1}$ and from $-$42 to $-$29~km~s$^{-1}$. Around the systemic
velocity of the cloud (i.~e., from $-$52 to $-$43~km~s$^{-1}$), the
distribution of the gas consists of compact and diffuse components. As
expected, the SMA data pick up the compact component well, while the diffuse
gas is only detected with the CSO single-dish observations. A velocity
gradient is seen in the channel maps, with a blue-shifted (from $-$69 to
$-$56~km~s$^{-1}$) component to the north-east of the W3(H$_2$O), and a
red-shifted component (from $-$40 to $-$32~km~s$^{-1}$) to the south-west. At
around $-$47 to $-$46~km~s$^{-1}$, we identify a flattened structure, without
diffuse emission, elongated in the southeast-northwest direction. This
velocity range is close to the self-absorption peak (see Section~3.2.2 and
Figure~3) for the HCN\,(3--2) line, suggesting that it traces a high-obscured
part of the core, therefore likely locates at the center of the dense
core. This structure is perpendicular to the high velocity blue and
red-shifted gas, which might indicate the presence of a disk (or toroid) and
outflow system. In general, the gas is mainly associated with W3(H$_2$O),
suggesting that HCN high velocity gas is most likely related to the presence
of W3(H$_2$O) rather than W3(OH). In addition to the high velocity gas
associated with W3(H$_2$O), emission from $-$44 to $-$40~km~s$^{-1}$ reveals a
north-south velocity gradient across W3(OH) which is consistent the CH$_3$OH
line as reported in Qin et al.\ (2015) and with the 6.7~GHz CH$_3$OH maser
observations of Menten et al.\ (1992).  

\begin{figure*} \includegraphics[width=12.81cm,angle=-90]{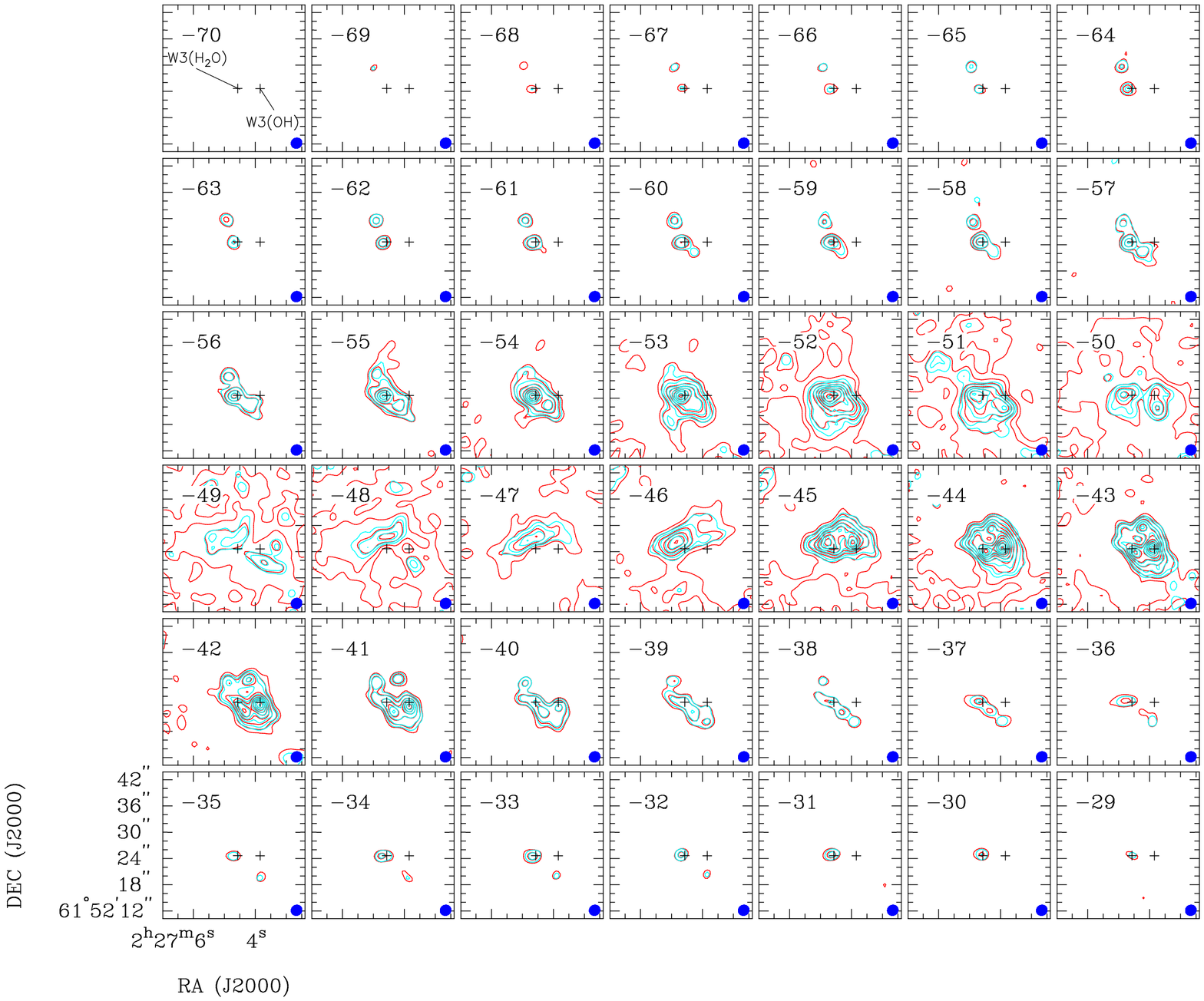}
 \caption{Channel maps of the HCN\,(3--2) transition from the combined SMA and
   CSO datasets (red contours) and from the SMA dataset alone (turquoise
   contours).  The contours are 5\%, 10\%, 20\%,30\%, 40\%, 50\%, 60\%, 70\%, 80\%, 90\% of the 
peak intensities, 12.7~Jy~beam$^{-1}$ for SMA+CSO and 12.2~Jy~beam$^{-1}$ for SM
A. The rms (1$\sigma$) noise levels for SMA+CSO and SMA data are 0.17 and 0.15~J
y~beam$^{-1}$, respectively. In each panel, the synthesized beam is shown in bot
tom-right corner, and the cross symbols indicates the peak positions of the W3(H
$_2$O) and W3(OH) continuum sources (Qin et al.\ 2015). }\label{}
\end{figure*}
 
\subsubsection{Spectra obtained with the SMA}

In Figure~3, we show the HCN spectrum obtained after combining the
SMA and CSO data (SMA+CSO, black line), and the HCN, HCO$^+$, HCN v$_2$=1
spectra as observed with the SMA (red, green, and blue lines) 
towards the W3(H$_2$O) and W3(OH) continuum peaks.  Both HCN and HCO$^+$
spectra towards the W3(H$_2$O) and W3(OH) show double peak line profiles while
weak HCN v$_2$=1 line is only seen towards the W3(H$_2$O). In principle, higher
energy transitions can only be exictated in the hot regions while lower energy
transitions present in the cold surrounding regions, but also in the hot
regions.  Qin et al. (2015) argued that non-detection of HCN v$_2$=1 towards the
W3(OH) is due to the fact that the gas seen in the W3(OH) comes from the outer
region of the UC H{\sc ii} region and inner portion of this hot core has been
disociated by the stars, and lower HCN abundance and gas temperature of envelope
make the HCN v$_2$=1 too weak and expected intensity is lower than the SMA
detection limit.  The broad line is seen in HCN but not in HCO$^+$ towards
the W3(H$_2$O) (see top panel of Figure 3),
suggesting that HCN is better tracing the molecular outflow of this
region. Both HCN and HCO$^{+}$ show an absorption dip at around
$-$47.8~km~s$^{-1}$, with the blue-shifted peak being stronger (`blue profile'
asymmetry). On the other side, the HCN v$_2$=1 line peaks at around
$\sim$$-$49~km~s$^{-1}$, i.~e., near the position of the absorption dip.  
The bottom panel of Figure~3 shows
the spectra towards the W3(OH) continuum peak position. Differently to the
case of W3(H$_2$O), the spectra are prominently red-shifted, i.~e., with the
red-shifted peak being stronger than the blue-shifted one. This kind of
profiles are indicative of gas expansion (e.~g., Aguti et al. 2007). 

 The combined SMA+CSO spectrum at the W3(H$_2$O) continuum peak (top panel of
 Figure 3) is only  slightly stronger than the SMA-only spectrum, indicating
 that even with only the SMA we are recovering most of the HCN emission, as
 expected for a tracer of compact and dense gas. The blue side of the SMA
 spectrum at the W3(OH) continuum peak is
much weaker than that of the SMA+CSO data. Previous observations of single dish and interferometers  (Mauersberger, Wilson \& Henkel 1988; Wilson et al. 1991, 1993; Wink et al. 1994) suggested that the W3(OH) complex consists of extended envelope with lower gas density of $\sim$10$^5$ cm$^{-3}$ and the compact core
 W3(H$_2$O) with gas density of $\sim$10$^7$ cm$^{-3}$. The critical density
of HCN (3--2) is at one order of $\sim$10$^7$ cm$^{-3}$ (at $\sim$50 K), so one
should see similar spectral profile at the W3(H$_2$O) position with the SMA and
SMA+CSO. This is also verified from the channel maps (see Figure 2), in which
compact clumps observed by the SMA and SMA+CSO have almost same morphology and
extended emission located outside of the compact clumps can only be seen by
the SMA+CSO data. As for the W3(OH), the weakness of blue side of the HCN
spectrum observed with the SMA should be caused by
the filtering effect of the interferometer in which lower density gas in front
of H{\sc ii} region  should show blue-shifted emission if the cloud
is in expansion, but can be recovered by single dish observations. 

The different line profiles may reflect different kinematcis between the
W3(H$_2$O) and W3(OH). In Section~3.2.2 and 3.3, we discuss the nature of these profiles.

\begin{figure*}

\includegraphics[width=0.6\textwidth,angle=-90]{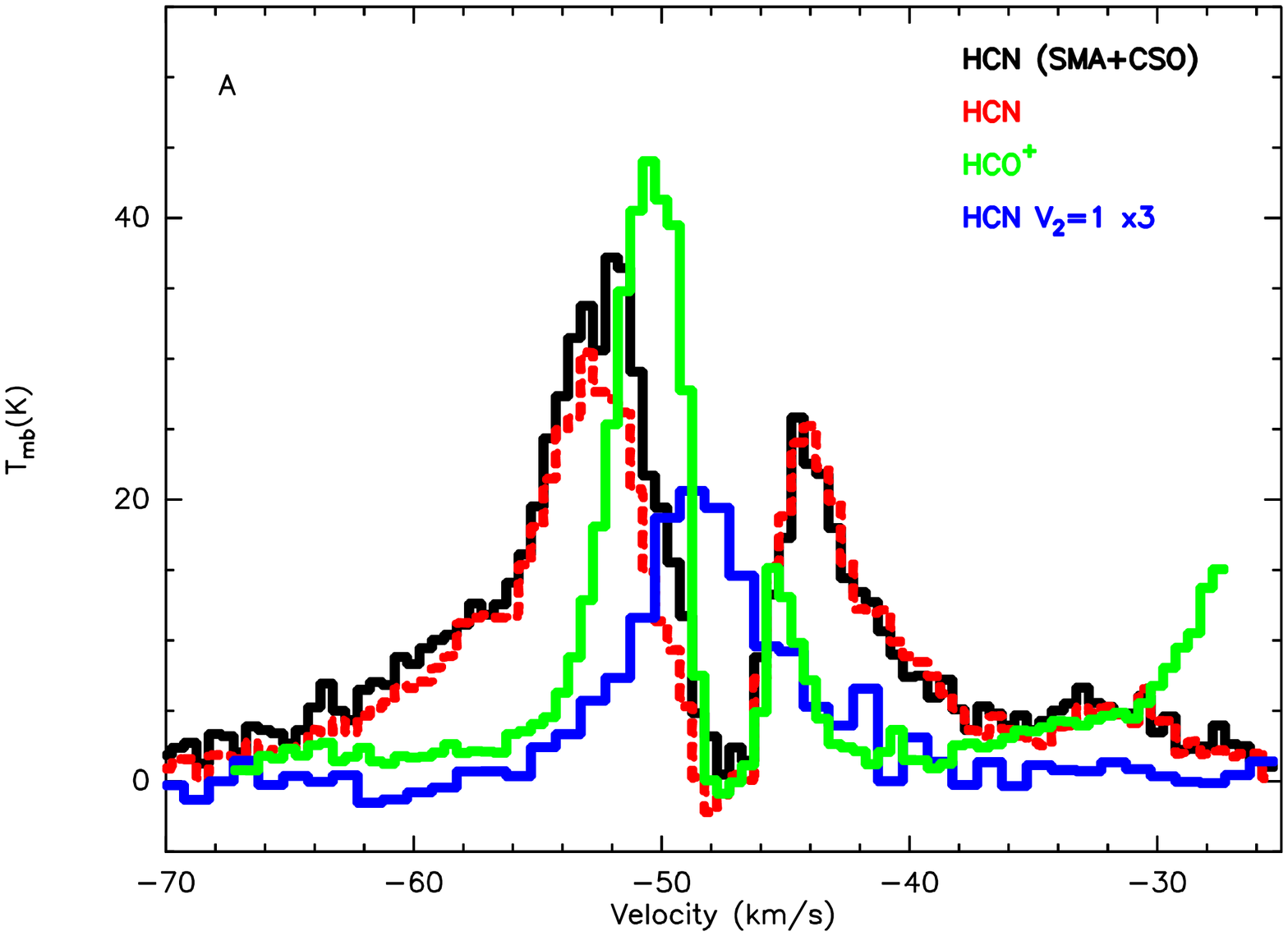}
\includegraphics[width=0.6\textwidth,angle=-90]{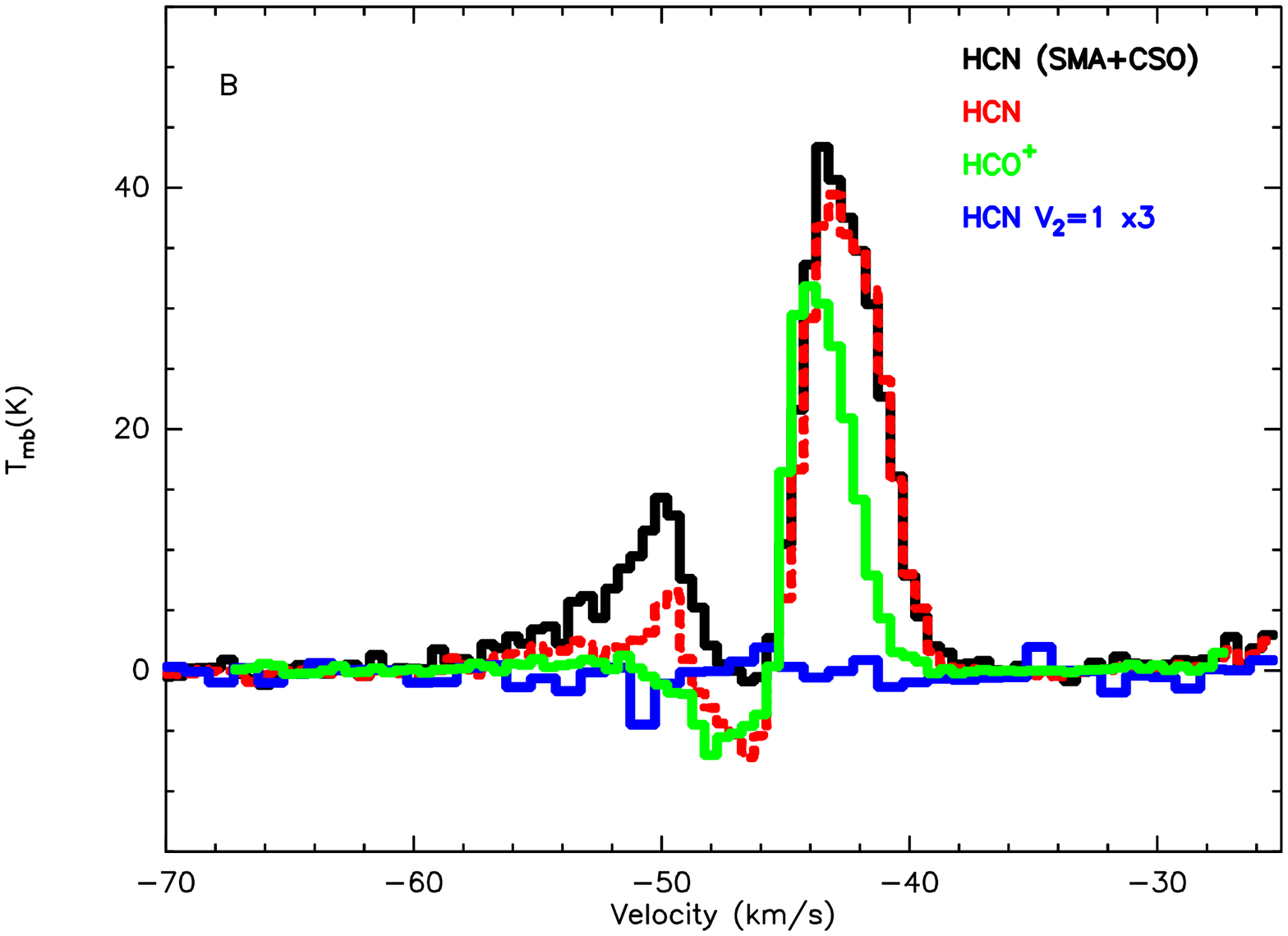}
 \caption{Top panel: Spectra extracted towards the W3(H$_2$O) continuum peak
   (see Qin et al.\ 2015) for HCN (SMA+CSO, black line), HCN (only SMA, red
   dashed line), HCO$^+$ (green line),  and HCN v$_2$=1 (blue line).  Bottom panel: Same as for the top panel towards the W3(OH) continuum peak. }\label{}
\end{figure*}

\subsection{\emph{Kinematics in W3(H$_2$O)}}
 Observations showed that molecular line emissions mainly arise from the
  W3(H$_2$O) core with higher gas density while absorption lines are obseved in
  the W3(OH) which have lower density (Mauersberger, Wilson \&
  Henkel 1988; Wilson et al. 1991, 1993, Wink et al. 1994). The two sources
  are spatially separated and  have different systemic velocities (Wilson
  et al. 1991; Qin et al. 2015), implying different physics and kinematics. At
  a large scale, the two sources share same envelope and the background
  sources will affect the observed optically thick line shapes in both line
  profile  and intensity if the background source has different kinematics. We
  think these will not obviously affect the observed line shape
  and intensity of the W3(H$_2$O) since higher density gas dominates
  kinematics of the compact core as seen from Figure 2 and Figure 3. 90\% of
  the molecular matter associated with the W3(OH) is located behind or to the
  side of the W3(OH) H{\sc ii} region (Wilson et al. 1991). Thus one should
  see `red asymmetry' line profile of HCN and HCO$^+$ with comparable read
  peak line intensity  observed by the SMA and SMA+CSO. An alternative
  explanation  of the line profile toward the W3(OH) is that W3(OH) may be
  located behind   W3(H$_2$O) and the blue-shifted HCN emission toward this 
source is absorbed by the dense envelope of the W3(H$_2$O), producing the
observed red asymmetry. If so, one should see similar velocity pattern as in the
W3(H$_2$O). But the blue-shifted peak velocities between the two cores are
different (see Figure 3), thus the observed `red  asymmetry' line profile  supports gas  expansion in the W3(OH) region. 

\subsubsection{Molecular outflows}

Previous observations of the CO\,(2--1) line show two large-scale bipolar
outflows driven by the W3(H$_2$O)-A and C sources (Zapata et
al.\ 2011). Compared to the CO transition, HCN\,(3--2) has a higher critical
density and should trace compact outflows close to the exciting
sources. Figure~4 presents the HCN velocity-integrated intensity map of the
SMA+CSO HCN line. Both the blue and red-shifted gas have three emission
clumps. The clumps located close to W3(H$_2$O) are likely associated with
W3(H$_2$O)-A, while the other clumps are located to the northeast and
southwest of the W3(H$_2$O)-C source, suggesting that they are tracing an
outflow powered by source C.  The outflow has a position angle of
54$^{\circ}$. Assuming that the HCN line wing emission is optically thin, an
excitation temperature of 50~K (as in Zapata et al.\ 2011), and an HCN
abundance of 9.6$\times$10$^{-9}$ (Qin et al.\ 2015), under local
thermodynamic equilibrium (LTE), we infer an outflow mass of
(2.6$\pm$0.3)~M$_{\odot}$ with some uncertainties due to the assumptions of
optically thin emission, excitation temperature and fractional abundance of
HCN. The estimated dynamic time of the outflow is
(5.6$\pm$0.2)$\times$10$^{3}$~yr, and the mass loss rate is
$\sim$(4.6$\pm$0.5)~$\times$10$^{-4}$~M$_{\odot}$~yr$^{-1}$.

\begin{figure*} \includegraphics[width=0.9\textwidth]{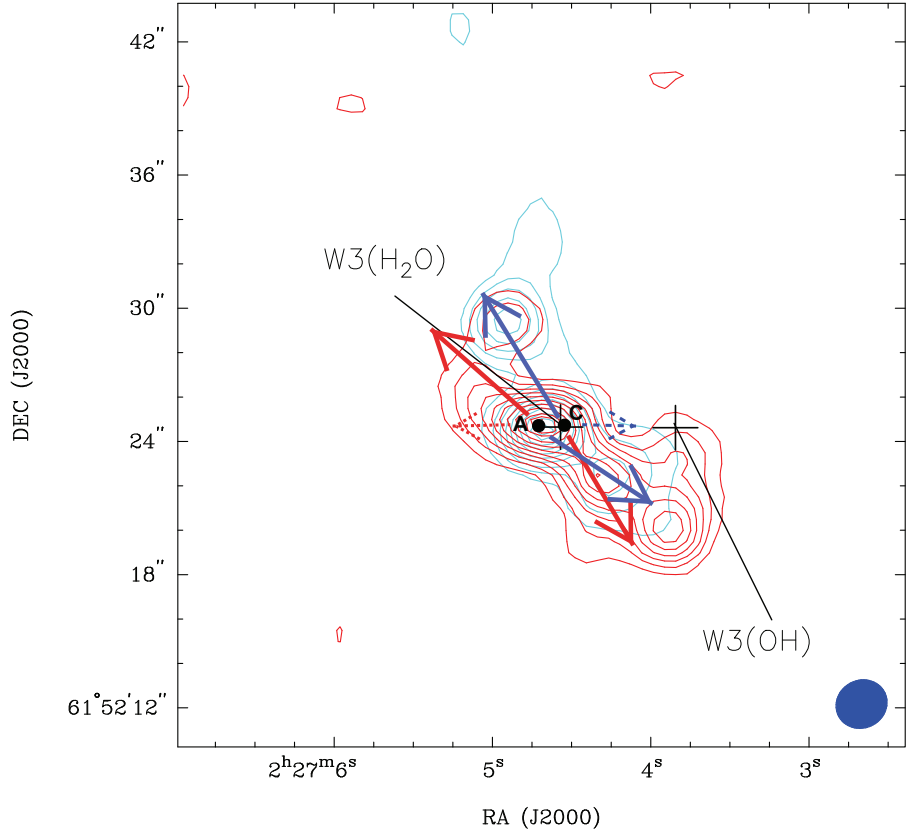}
 \caption{HCN outflow image from the CSO+SMA data.  The blue-shifted emission (blue contours) show the emission integrated from $-$69 to $-$56~km~s$^{-1}$, while the red-shifted (red contours) show the emission from $-$40 to $-$32~km~s$^{-1}$. The contours levels of both the blue (peak value of 20.6~Jy~beam$^{-1}$~km~s$^{-1}$) and red lobes (14.8~Jy~beam$^{-1}$~km~s$^{-1}$) range from 10\% to 100\% of the peak emission. The crosses indicate the peak positions of the W3(H$_2$O) and W3(OH) continuum (Qin et al.\ 2015). The filled circles denote the W3(H$_2$O)-A and C source positions. The synthesized beam is shown in bottom-right corner. The blue and red arrows denote the orientation of CO molecular outflows (Zapata et al.\ 2011).  The dashed arrows indicate the orientation of non-thermal jets and water maser outflows (Reid et al.\ 1995; Alcolea et al.\ 1993) }\label{}
\end{figure*}

\subsubsection{Infall motions}

The HCN and HCO$^+$ SMA spectra toward the W3(H$_2$O) continumm peak position
show an absorption dip at around the systemic velocity ($-$47.8~km~s$^{-1}$,
see Figure~3-top).  This absorption dip is also visible in the single-dish
data (see Figure~1 of Wu \& Evans 2003), indicating that it is not caused by
interferometric spatial filtering. Another possible explanation is the
presence of two different velocity components, but this is ruled out when
examining the higher energy transition of HCN v$_2$=1, which is expected to be
optically thin and shows a single peak at the position of the absorption dip
(see Figure~3-top). Therefore, the profile seen in the HCN and HCO$^+$ lines
is likely due to these lines being optically thick. Moreover, the blue-shifted
peak is stronger than the red-shifted one, showing  `blue profile' asymmetry. The classical  `blue asymmetry' says that the optically thin will peak in the
middle of absorption dip of the optically thick line if the emission is from a
single source. But in the W3(H$_{2}$O) case, higher angular observations of
continuum and lines have resolved W3(H$_{2}$O) into A and C subcomponents with
different systemic velocities, in which sources A and C have systemic
velocities of --51.4 and --48.6 km~s$^{-1}$ (Chen et al. 2006), and --50.5 and
--46.5 km s$^{-1}$ (Zapata et al. 2011) respectively. The HCN v$_2$=1 line
presents east-west velocity gradient (Qin et al. 2015) and the emission may
contribute from both sources A and C. Therefore the peak velocity of
 HCN v$_2$=1 does not peak at the HCN absorption dip. Our `two-layer model'
gave a medium value of --47.8 km s$^{-1}$ (see next paragraph) between
--48.6 and --46.5 km s$^{-1}$ of C component, indicating that gas
in falling onto the W3(H$_2$O)-C. Following Mardones et al. (1997),
we use the asymmetry parameter $\delta v$ --- the velocity difference between
the peaks of the optically thick line ($v_{\rm thick}$) and the optically thin
($v_{\rm thin}$) lines --- to quantify the blue asymmetry of the HCN and
HCO$^+$ lines in W3(H$_2$O). The parameter is defined as $\delta v$=($v_{\rm
  thick}-v_{\rm thin})$/$\Delta$$v_{\rm thin}$, with $\Delta$$v_{\rm thin}$
being the line width of the optically thin HCN v$_2$=1. An statistically
significant excess of blue asymmetric line profiles with $\delta v$$<$$-0.25$
indicate that the molecular gas is infalling onto the cores (e.~g., Mardones
et al.\ 1997; Wu \& Evans 2003; Fuller, Williams \& Sridharan 2005; Wu et
al.\ 2007). Gaussian fitting is made to the HCN\,(3--2), HCO$^+$\,(3--2) and
HCN v$_2$=1 lines. The resulting $\delta v$ for HCN\,(3--2) and
HCO$^+$\,(3--2) are $-0.65$ and $-0.38$, respectively. These values indicate
gas infalling towards the W3(H$_2$O) continuum core.

 Although the asymmetry parameters devived from both HCN\,(3--2) and
  HCO$^+$\,(3--2) lines suggest gas infalling motion, the HCN spectrum in the W3(H$_2$O) may be confused by the outflows as seen from Figure 2 and Figure 3. In the case of HCO$^+$ (1--0), the interferometer see 40\% of the single dish flux
(Wink et al. 1994).  HCO$^+$ (3-2) has larger
  critical denisty compared to HCO$^+$ (1--0) should recover more flux of the
single dish observations. In order to quantify the infall rate, we modeled
the `blue profile' HCO$^+$ spectrum using a two-layer model (see Appendix, and
Myers et al. 1996; Di Francesco et al. 2001). The model fits the observed
spectrum by setting the background temperature of the continuum source, the
excitation temperature, the beam filling factor, the systemic velocity, the
infall velocity, the velocity dispersion, and the peak optical depth. The best
fit to the HCO$^+$ spectrum (see Figure~5) results in
an infall velocity ($V_{in}$) of  2.7$\pm$0.3~km~s$^{-1}$ and a velocity
dispersion of 0.8$\pm$0.1~km~s$^{-1}$. Similar to the work done by Pineda et
al. (2012), we assume that the gas is in a spherically symmetric free fall
onto the W3(H$_2$O) core.  Therefore, we can estimate the mass infall rate
($\dot{M}=3V_{in}^{3}/2G$) to be (2.3$\pm$0.7)$\times$10$^{-3}$~M$_{\odot}$~yr$^{-1}$. 
\begin{figure*}
\includegraphics[width=0.45\textwidth]{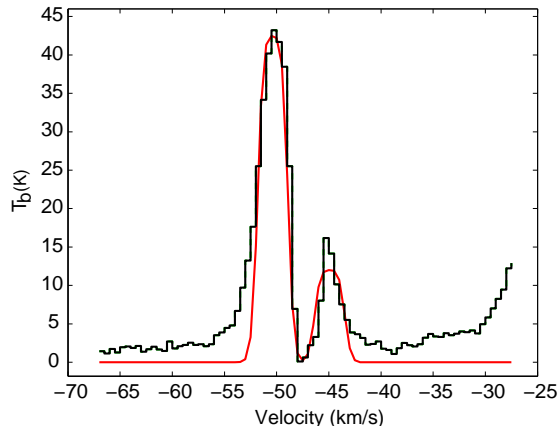}
\caption{ Observed HCO$^+$  (black line) and two-layer model (red) spectra towards the W3(H$_2$O) continuum peak position (see Qin et al.\ 2015).  }
\end{figure*}
\subsection{\emph{Expansion in W3(OH)}}

 Figure~6 (left panel) shows the HCN spectrum (CSO+SMA) at the W3(OH)
 continuum peak position. It shows an absorption dip at around the systemic
 velocity ($-$46.4~km~s$^{-1}$) with the red-shifted peak being stronger than
 the blue-shifted one (`red profile'), suggestive of expansion. In the right
 panel, we show the HCO$^+$ spectrum, which reveals a P-cygni profile. The
 two-layer model (Myers et al.\ 1996) was used to fit the observed `red
 profile' and determine the expansion velocity. Note that we set same filling
 factor and background temperature for the HCN and HCO$^+$ lines in the
 model. The mean expansion velocities are estimated to be
 3.8$\pm$0.4~km~s$^{-1}$ and 1.6$\pm$0.2~km~s$^{-1}$ from the HCN (CSO+SMA)
 and HCO$^+$ lines (SMA), respectively.  The critical denisty of HCN (3--2)
   is one order of magnitude higher than that of HCO$^+$ and then one expects 
  HCN to be more opaque in denser regions (Wu \& Evans 2003; Reiter et
  al. 2011). HCN and HCO$^+$ are likely tracing different layers
 of the cloud, which lead to different expansion velocities. More quantitative
 analyses by use of radiative transfer model such as RATRAN (Hogerheiheijde \&
 van der Tak 2000) are needed. As done by Rolffs et al. (2010),
 modeling multiple transitions of HCN or HCO$^+$ spanning a wide energy levels
 with different opacities can constrain velocity field, abundance and density
 profiles, and kinematics well. Currently the density and temperature profiles
 are only obtained towards the W3(H$_2$O) core (Chen et
 al. 2006). Observations of multilevel transitions of the HCN or HCO$+$ are
 needed to obtain the physical and kinematical structures toward the W3(OH) complex in the future. 
\begin{figure*}

\begin{tabular}{c c}
\includegraphics[width=0.45\textwidth]{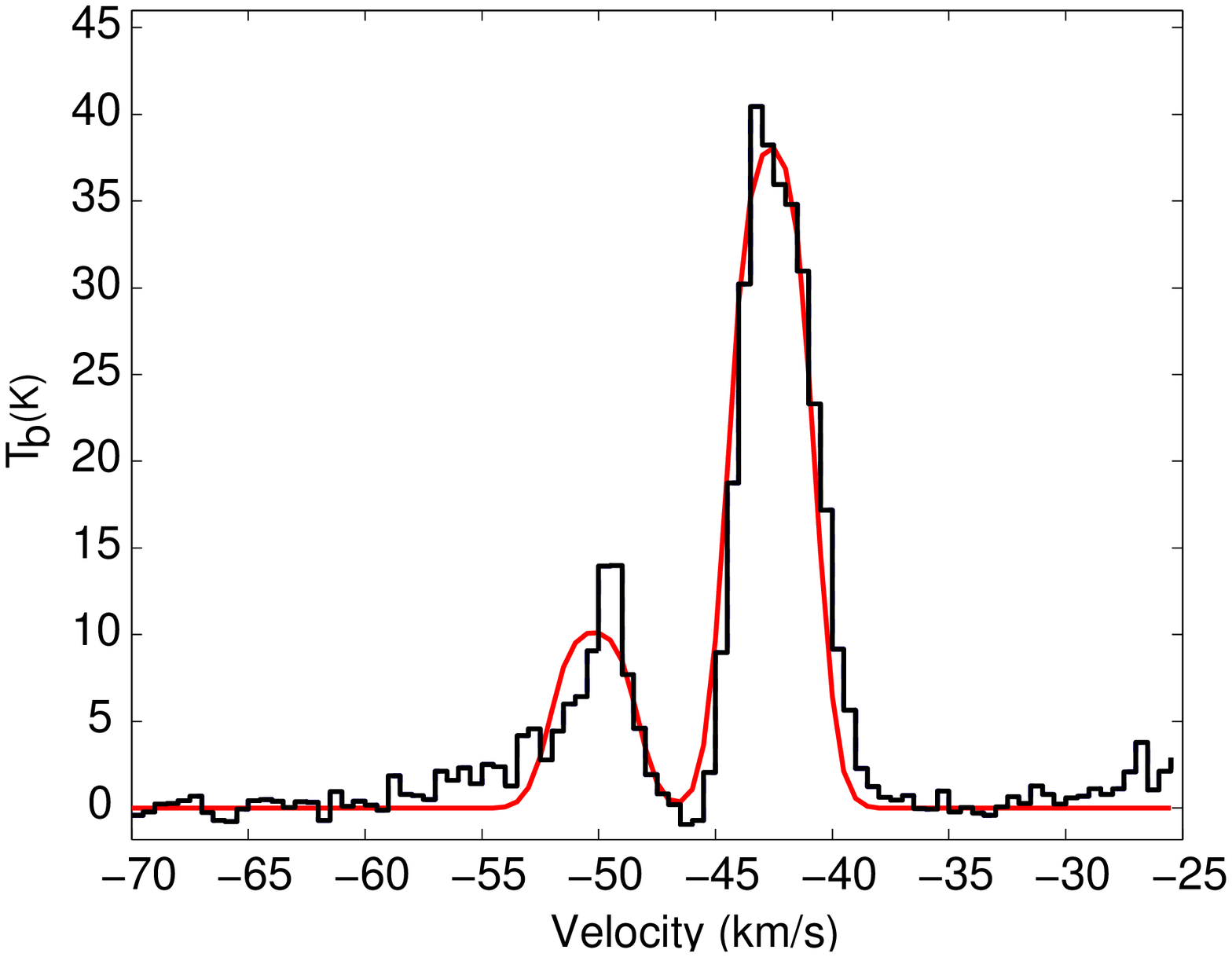} &
\includegraphics[width=0.45\textwidth]{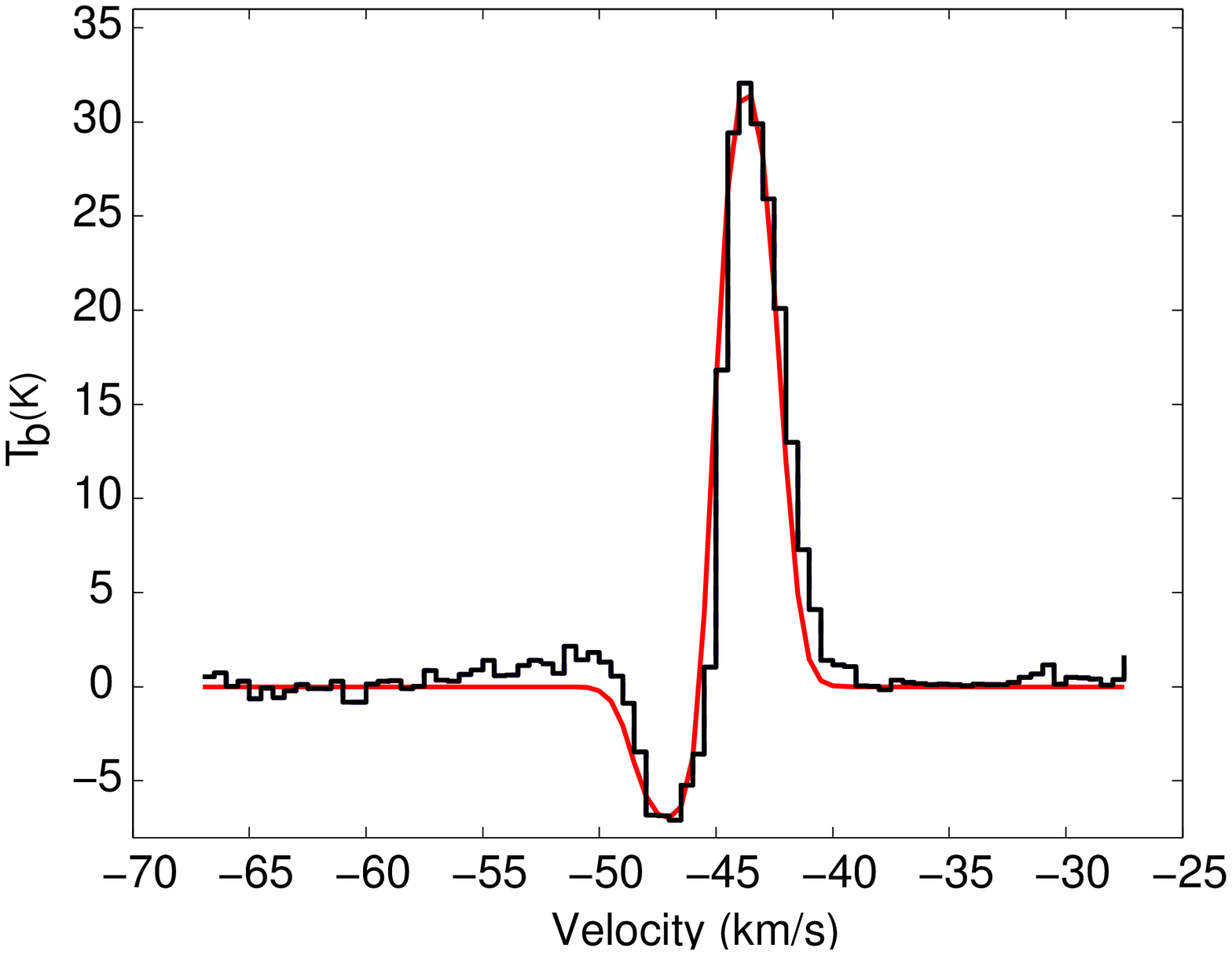} \\
\end{tabular}
\caption{HCN (left panel, from CSO+SMA datasets) and HCO$^+$ (right panel, from SMA dataset) spectra towards the W3(OH) continuum peak position (see Qin et al.\ 2015).  Left panel: observed (black line) and two-layer model (red) HCN spectra.  Right panel: observed (black) and two-layer model (red) HCO$^+$ spectra.}

\end{figure*}

In Figure~7, we present the flux ratio map of the \emph{Spitzer}/IRAC
4.5~$\mu$m to 3.6~$\mu$m emission (in color scale) overlaid with the SMA
1.1~mm continuum emission (in contours) towards the W3(OH) complex.  The
observations and analyses suggested that the morphologies observed at
3.6~$\mu$m  and 4.5 $\mu$m are similar to each other and to H$_2$ 1--0 S(1)
line, and thus the emissions of the 4.5~$\mu$m  and 3.6~$\mu$m  bands include
bright H$_2$ lines (e.~g., Cyganowski et al.\ 2009; Takami et al. 2010)
compared to the other bands of the \emph{Spitzer}/IRAC instrument. It is thought that a 4.5~$\mu$m to 3.6~$\mu$m
flux ratio larger than 1.5 traces shocked gas (e.~g., Takami et al.\ 2010). As
shown in the figure 7, both the 4.5 and 3.6~$\mu$m images are saturated within
the 1.1~mm continuum emission (the region encompassed by the pink circle), and a reliable ratio image can not be
obtained. However, the ratio can be determined in the non-saturated
pixels. Values of the ratio $>1.5$ are found in an area of diameter
$\sim$50$^{\prime\prime}$. The ratio map shows a gradient distribution
decreasing towards the outer regions, with the largest values found  around the W3(OH) continuum source. This result might indicate large-scale shocks in the region. Interestingly, pixels with ratios larger than 3 are found in three ring-like structures surrounding the W3(OH) continuum peak, mainly distributed to the south. This can be interpreted as the ambient gas not being uniform, if the shocked material is due to the expansion of the H{\sc ii} region.
\begin{figure*}

\includegraphics[width=1.2\columnwidth,angle=-90]{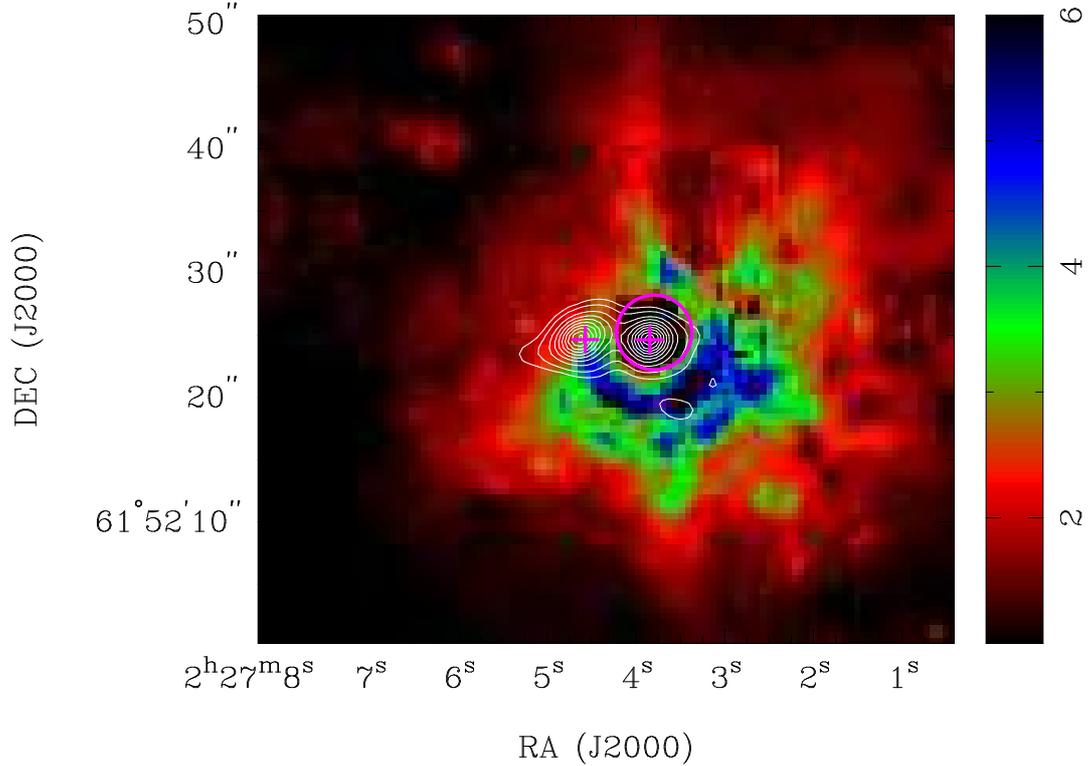}
\caption{\emph{Spitzer}/IRAC 4.5~$\mu$m-to-3.6~$\mu$m ratio image (colour
  scale) overlaid with the SMA 1.1~mm continuum map (contours). The crosses
  indicate the peak positions of the W3(H$_2$O) and W3(OH) continuum
  sources. Note that the \emph{Spitzer}/IRAC 4.5~$\mu$m and 3.6~$\mu$m images
  are saturated at the position of W3(OH) continuum as encompassed by the pink
  circle symbol. }

\end{figure*}

\section{Discussion}
\subsection{\emph{Outflows and infall}}

At least two CO molecular outflows are identified in the W3(OH) complex,
likely driven by the sources W3(H$_2$O)-A and C (Zapata et al.\ 2011). In
comparison, the outflows that we have detected in the HCN\,(3--2) transition
show a smaller spatial extension and outflow mass, which could be explained by
the fact that the CO traces large-scale and extended outflowing material. The
directions of the outflows determined for both HCN and CO outflows are
consistent (see arrows in Figure~4), associated with W3(H$_2$O)-C, and show
multiple knots in the red and blue-shifted lobes. So far, only a few massive
outflows associated with high-mass star forming regions have been observed to
show a discrete knotty structure (e.~g., Qiu \& Zhang 2009; S\'anchez-Monge et al.\ 2014). One possible interpretation is that
the HCN outflows in our observations are made of swept-up ambient gas
originated from an episodic and precessing mass-loss process (Arce \& Goodman
2001).  Non-thermal radio jets and water maser outflows along the east-west
direction (PA$=90^{\circ}$) revealed by the VLA and VLBI observations (e.~g.,
Wilner et al. 1999), two CO molecular outflows (with PA$\sim$$40^{\circ}$ and
$\sim$15$^{\circ}$; Zapata et al.\ 2011) and our HCN outflows
(PA$\sim$$54^{\circ}$) suggest that different tracers trace different
kinematics of the outflows, providing another evidence for jet precession in
the W3(H$_2$O). A binary system (and disk candidates) have been reported in
the W3(H$_2$O) (e.~g., Wyrowski et al.\ 1999; Chen et al.\ 2006; Zapata et
al.\ 2011). It has been proposed that episodic and precessing outflows might
be related to the presence of a binary system (e.~g.\ Arce \& Goodman
2001). Therefore, episodic and precessing outflows in the W3(H$_2$O) might be
related to the presence of a binary system. The lack of an H{\sc ii} region in
W3(H$_2$O), the presence of strong dust emission and the derived short outflow
timescale of (5.6$\pm$0.2)$\times$10$^{3}$~yr, compared with other high-mass
molecular outflows (e.~g., Beuther et al.\ 2002; Wu et al.\ 2004; Zhang et
al.\ 2005; S\'anchez-Monge et al.\ 2013b), suggest that the W3(H$_2$O) could
be in an early evolutionary stage of the high-mass star formation process.

Spectral infall signatures are seen in W3(H$_2$O), and an unusually large
spherical mass infall rate of
(2.3$\pm$0.7)$\times$10$^{-3}$~M$_{\odot}$~yr$^{-1}$ is derived. Recent
observations have revealed high mass accretion rates
(10$^{-3}$--10$^{-2}$~M$_{\odot}$~yr$^{-1}$) in other high-mass star forming
regions (e.~g., Sandell, Goss \& Wright 2005; Garay et al.\ 2007; Zapata et
al.\ 2008; Wu et al.\ 2009; Rolffs et al.\ 2010; Liu et al.\ 2011a,
2013). This work predicts comparable accretion rates, suggesting high-mass
stars are forming in W3(H$_2$O). Adopting an H$_2$ number density of
10$^{8}$~cm$^{-3}$ (Chen et al.\ 2006), and assuming that the collapse is
uniform, pressure-free, and non-rotating, the free-fall time
($t_{ff}=\sqrt{3\pi/(32G\rho)}=5\times10^{7}/\sqrt{n(H_2)}$) is estimated to
be 5$\times$10$^3$~yr, which is consistent with the outflow time scale. The
assumptions are reasonable since the derived density profile of W3(H$_2$O)
support free fall collapse (Chen et al.\ 2006).  The infall velocity derived
from a two-layer collapse model fit to the HCO$^+$\,(3--2) line is
2.7$\pm$0.3~km~s$^{-1}$. Assuming that the velocity dispersion
(0.8$\pm$0.1~km~s$^{-1}$) is caused by thermal motions, the large infall
velocity suggests that collapse is proceeding in a supersonic way. Chen et
al.\ (2006) derived the W3(H$_2$O)-A and C systemic velocities of $-$51.4 and
$-$48.6~km~s$^{-1}$, respectively, while Zapata et al.\ (2011) derive systemic
velocities of $-$50.5 and $-$46.5~km~s$^{-1}$ for the A and C sources.  Our two-layer model fitting to the  HCO$^+$ line results in a systemic velocity of $-$47.8~km~s$^{-1}$, which seems to indicate that the gas is infalling onto the W3(H$_2$O)-C.

\subsection{\emph{ Expansion}}

 Based on the relative proper motions of the OH maser spots observed at
 different epochs towards W3(OH), Bloemhof, Reid \& Moran (1992) proposed gas
 expansion from the UC H{\sc ii} region to explain the observed pattern of
 motions. Furthermore, a direct measurement of expansion of the W3(OH) was
 made through the different difference maps obtained by the VLA at three
 epochs (Kawamura \& Masson 1998). They suggest that the shell-like H{\sc ii}
 region is slowly expanding with a typical speed of 3--5~km~s$^{-1}$ and
 derive a dynamical age of $\sim$2300~yr.

The HCN red profile and HCO$^+$ P-cygni profile towards the W3(OH) continuum peak position suggest that the gas in the envelope is expanding (see Figure~6). The two-layer model results in mean expansion velocities of 3.8$\pm$0.4~km~s$^{-1}$ and 1.6$\pm$0.2~km~s$^{-1}$, respectively. Considering the different critical densities of the two transitions, this suggests that the expansion of the envelope is decelerating. Flux ratio map of \emph{Spitzer}/IRAC 4.5~$\mu$m and 3.6~$\mu$m images shows a radial distribution from inside to outside, with the largest ratios found around the W3(OH) continuum source, likely pinpointing W3(OH) as the expansion center. Ratios larger than 1.5 indicate shocked gas, likely produced by the interaction of the expanding H{\sc ii} region and the molecular cloud. At the distance of W3(OH) (2.2~kpc; Hachisuka et al.\ 2006), the shock radius of 25$^{\prime\prime}$ (see Figure~7) and the expanding velocity of 3.8$\pm$0.4~km~s$^{-1}$ (derived from HCN line) result in an expanding timescale of about (6.4$\pm$0.7)$\times$10$^4$~yr. This value is larger than the timescale of 2300~yr derived by Kawamura \& Masson (1998), but lower than the typical age of 10$^5$~yr for UCH{\sc ii} regions. We do think that the expansion timescale of W3(OH) as derived in this work is reasonable, since the timescale of infall and outflow motions in the W3(H$_2$O) is approximately 5$\times$10$^3$~yr, and W3(H$_2$O) is likely at an earlier evolutionary stage compared to W3(OH).

\subsection{\emph{Dynamical state of the W3(OH) complex}}

 The W3(OH) complex is an nearby star formation region harboring two sites of
 high-mass star formation, W3(H$_2$O) and W3(OH), with different
 properties. W3(H$_2$O) is rich in water masers and organic molecules, and at
 earlier evolutionary stage compared to W3(OH), which is associated with an
 UCH{\sc ii} region and is rich in OH masers. Previous observations of OH
 masers and NH$_3$ absorption features suggest a collapsing outer envelope
 outside the UCH{\sc ii} region. Our CSO grid spectra of the HCN\,(3--2) line
 (see Figure~1) show a `blue profile' in a large area suggesting that collapse
 is not happening only around the UCH{\sc ii} region but it is a global
 collapse in the W3(OH) complex.

At small scales, the two sources have different kinematics, based on both
previous studies and the observations presented in this work. A small-scale
non-thermal radio jet and water maser outflows along the east-west direction
together with large scale CO outflows and a rotating disk candidate have been
reported towards in W3(H$_2$O) in previous works. Complementary to this, our
HCN and HCO$^+$ line results show infall and outflow motions, and a flattened
structure perpendicular to the direction of the outflows. At the same time,
relative proper motions of OH maser spots and multi-epoch radio continuum maps
suggest the presence of expansion motions originated from the W3(OH) UCH{\sc
  ii} region. Our molecular line observations, combined with
\emph{Spitzer}/IRAC images confirms the expansion interpretation with W3(OH)
being located at the center. From the expansion velocity derived with the
two-layer model fit and the size of the shocked gas as revealed by the IRAC
images, we determine an expansion timescale of (6.4$\pm$0.7)$\times$10$^4$~yr.

Based on the above statements, we present a sketch of the dynamical state of
the W3(OH) complex in Figure~8. We describe the picture in the following. The
W3(OH) complex harbors W3(H$_2$O) and W3(OH), with W3(H$_2$O) being in an
earlier evolutionary stage of massive star formation compared to W3(OH), who
has already developed an H{\sc ii} region. Far from the forming stars, the
outer envelope of the W3(OH) complex is collapsing inwards. Inside the
complex, W3(H$_2$O) is still undergoing collapse and is associated with
collimated outflowing material. On the other side the H{\sc ii} region of
W3(OH) is expanding and driving a shock front into the ambient material, which
leads to the observed neutral gas expansion and shock. 
\begin{figure*}

\includegraphics[width=1.0\textwidth]{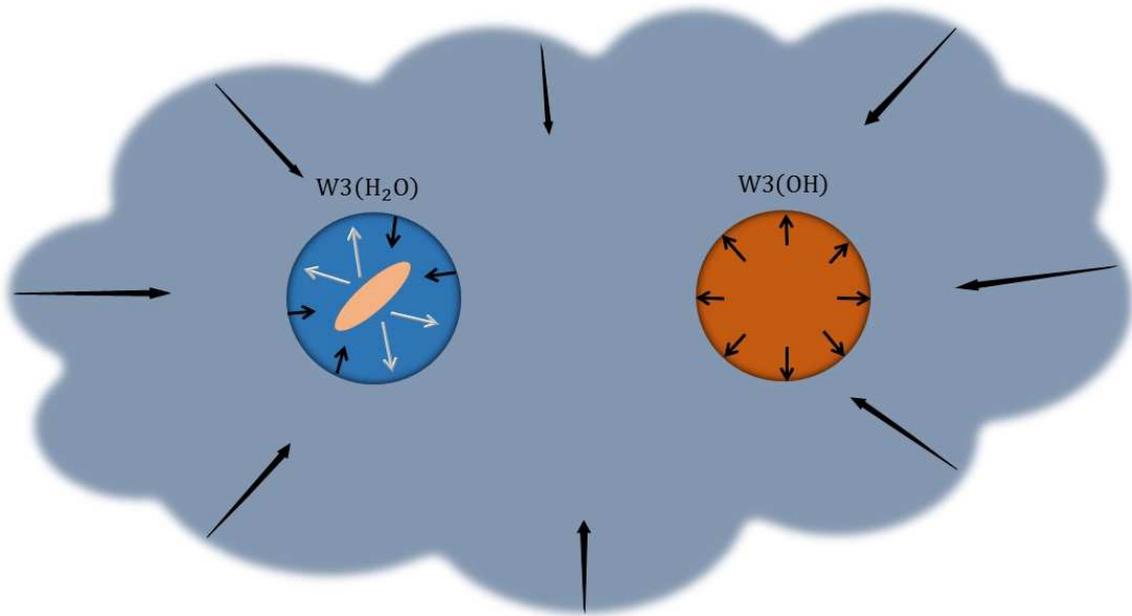}
\caption{Sketch of dynamical state of the W3(OH) star forming complex. The molecular cloud harboring the W3(OH) and W3(H$_2$O) regions is globally collapsing. Gas expansion is seen towards W3(OH) due to the expansion of the H{\sc ii} region. At the position of W3(H$_2$O) there is infall (black arrows), outflows (white arrows) and a possible rotating disk (yellow ellipse).}

\end{figure*}
\section{Summary}

 We have carried out SMA and CSO observations of different molecular tracers (HCN and HCO$^+$ lines) towards the W3(OH) star forming complex. The results are summarized in the following.

\begin{itemize}
\item The CSO grid spectral observations of HCN show a blue asymmetric profile over a large area covering the whole star forming complex, which suggests that the outer envelope of the W3(OH) complex is undergoing global collapse.

\item The SMA (and SMA+CSO) observations reveal blue asymmetric line profiles
  for both HCN and HCO$^+$ spectral lines towards the W3(H$_2$O) dust
  continuum core. A two-layer collapse model is used to fit the HCO$^+$
    line shape. The derived mass accretion rate is
  (2.3$\pm$0.7)$\times$10$^{-3}$~M$_{\odot}$~yr$^{-1}$. The free fall timescale is
  estimated to be 5$\times$10$^{3}$~yr. 

\item Bipolar outflows, identified in the HCN images, are likely driven by the source C in W3(H$_2$O). The molecular outflows show a knotty structure, which might be caused by an episodic and precessing process, likely related to the presence of a binary system in W3(H$_2$O). A mass loss rate of (4.6$\pm$0.7)$\times$10$^{-4}$~M$_{\odot}$~yr$^{-1}$ is derived, larger than the typical values found in low-mass molecular outflows. The dynamic time scale of the outflows is around (5.6$\pm$0.2)$\times$10$^{3}$~yr, consistent with the free fall timescale. The large accretion and mass loss rates, confirm the formation of high-mass stars in the W3(H$_2$O) region.

\item The red asymmetric profile of the HCN line together with the HCO$^+$ P-cygni profile at the W3(OH) continuum peak, indicate that the envelope around this source is expanding. A two-layer model fit to the spectra results in expansion timescales of (6.4$\pm$0.7)$\times$10$^{3}$~yr. The \emph{Spitzer}/IRAC 4.5~$\mu$m-to-3.6~$\mu$m flux ratio image is also consistent with the envelope around W3(OH) being in expansion, and shows a gradient distribution from inside to outside with largest values (i.~e., more evidence of shock, likely related to a faster expansion) around the W3(OH) continuum source. We conclude that the W3(OH) H{\sc ii} region is expanding and interacting with the ambient gas, and the shocked neutral gas is decelerating expanding.

\end{itemize}

In summary, the W3(OH) complex is undergoing global collapse. Inside the complex, the two objects W3(H$_2$O) and W3(OH) are at different evolutionary stage of massive star formation. Infall and outflow motions are observed at an earlier stage in the W3(H$_2$O), while the W3(OH) H{\sc ii} region is expanding, which leads to observed neutral gas  expansion and shock.

\section*{Acknowledgements}

 We thank the anonymous referee, and editor Morgan Hollis for their
 constructive comments on the paper. This work has been supported by the National Natural Science Foundation of
 China under grant Nos. 11373026, 11373009, 11433004, 11433008, U1331116, and
 the National Basic Research Program of China (973 Program) under grant
 No. 2012CB821800, by Top Talents Program of Yunnan Province (2015HA030) and Midwest universities comprehensive strength promotion project (XT412001, Yunnan university), by the Deutsche Forschungsgemeinschaft, DFG through project number SFB956.

\clearpage
\newpage
\appendix

{\bf Appendix: Two-layer model}

 The two-layer model (Myers et al.\ 1996; Di Francesco et al.\ 2001) is aimed at fitting the observed spectrum by setting the radiation temperatures of the continuum source $J_{\rm c}$, the excitation temperature of the "rear" and "front" layers $J_{\rm r}$ and $J_{\rm f}$, the cosmic background radiation $J_{\rm  b}$, the beam filling factor $\Phi$, the systemic velocity $V_{\rm LSR}$, the kinematic velocity $V_{\rm k}$, the velocity dispersion $\sigma$ and the peak optical depth $\tau_0$. The predicted line brightness temperature at velocity $V$ is given by
\begin{equation}
\Delta T_{\rm B}=(J_{\rm f}-J_{\rm cr})[1-{\rm exp}(-\tau_{\rm f})]+(1-\Phi)(J_{\rm r}-J_{\rm b})[1-{\rm exp}(-\tau_{\rm r}-\tau_{\rm f}],
\end{equation}
where
\begin{equation}
J_{\rm cr}=\Phi J_{\rm c}+(1-\Phi J_{\rm r}),
\end{equation}
\begin{equation}
\tau_{\rm f}=\tau_0 {\rm exp}[\frac{-(V-V_{\rm LSR}-V_{\rm k})^2}{2\sigma^2}],
\end{equation}
\begin{equation}
\tau_{\rm r}=\tau_0 {\rm exp}[\frac{-(V-V_{\rm LSR}+V_{\rm k})^2}{2\sigma^2}],
\end{equation}
and the radiation temperature is expressed as $J_{\rm x}=\frac{T_0}{\rm exp(T_0/T)-1}$.

In this model the kinematic velocity is given by
\begin{equation}
V_{\rm k}=\frac{\sigma^2}{V_{\rm R}-V_{\rm B}}{\rm ln}\frac{1+{\rm exp}(T_{\rm BD}/T_{\rm D})}{1+{\rm exp}(T_{\rm RD}/T_{\rm D})},
\end{equation}
where $T_{\rm D}$ is the brightness temperature of absorption dip, and $T_{\rm BD}$ and $T_{\rm RD}$ are the intensity of the blue and red peaks above the dip, respectively. Positive $V_{\rm k}$ corresponds to gas infalling onto the continuum core, while negative $V_{\rm k}$ indicates gas expansion (e.~g., Myers et al.\ 1996; Aguti et al.\ 2007).

We applied the `lsqcurvefit' method in \texttt{matlab} to fit the spectrum. We
fitted the HCO$^{+}$ spectra first by setting all the parameters free. The
line profile is mainly determined by $V_{\rm LSR}$, $\sigma$ and $V_{\rm k}$,
while the intensity of the two peaks is mainly determined by $\tau_0$, $\Phi$,
$J_{\rm c}$, $J_{\rm f}$ and $J_{\rm r}$. In the model, we set the ranges for
$\tau_0$ from 0.1 to 10, for the filling factor from 0 to 1, $J_{\rm c}$,
$J_{\rm f}$ and $J_{\rm r}$ from 10 to 100~K, $V_{\rm LSR}$ from $-$50 to
$-$43~km~s$^{-1}$, $\sigma$ from 0.1 to 3~km~s$^{-1}$, $V_{\rm k}$ from 0.1 to
6~km~s$^{-1}$. For each set of initial values, `lsqcurvefit' will be converged
to a solution and try to give the best estimated values. We find that
$\tau_0$, $\Phi$, $J_{\rm c}$, $J_{\rm f}$ and $J_{\rm r}$ are very sensitive
to initial values and cannot be determined accurately due to
degeneracy. Contrary to that, $V_{\rm LSR}$, $\sigma$ and $V_{\rm k}$ are much
less sensitive to the initial input values. When we change the initial values,
the output for $V_{\rm LSR}$, $\sigma$ and $V_{\rm k}$ only varies by less
than 10\%. Thus, the error of the inferred infall speed is expected to be
smaller than 10\%.
\label{lastpage}

\end{document}